\documentclass[12pt] {article}
\usepackage{latexsym}{\rm }
\usepackage[dvips]{graphics,color}
\usepackage{graphpap}

\newcommand{\beq}{\begin{equation}}
\newcommand{\feq}[1]{\label{#1} \end{equation}}
\newcommand{\beqr}{\begin{eqnarray}}
\newcommand{\feqr}{\end{eqnarray}}
\def\non{\nonumber}
\def\noi{\noindent}

\newcommand{\rf}[1]{(\ref{#1})}
\def\pr{^{\prime}}
\definecolor{red}{rgb}{1,0,0}
\def\pr#1#2#3{Phys. Rep. {\bf{#1}} (#2) #3}

\def\pre#1#2#3{Phys. Rev. {\bf{E#1}} (#2) #3}

\def\rmp#1#2#3{Rev. Mod. Phys. {\bf{#1}} (#2) #3}

\def\pra#1#2#3{Phys. Rev. {\bf{A#1}} (#2) #3}
\def\jsp#1#2#3{J. Stat. Phys. {\bf{#1}} (#2) #3}
\def\chaos#1#2#3{Chaos {\bf{#1}} (#2) #3}

\def\jmp#1#2#3{J. Math. Phys. {\bf{#1}} (#2) #3}

\def\physica#1#2#3{Physica {\bf{A#1}} (#2) #3}

\def\acp#1#2#3{Adv. Chem. Phys. {\bf{A#1}} (#2) #3}
\def\rmp#1#2#3{Rev. Mod. Phys. {\bf{A#1}} (#2) #3}
\def\tams#1#2#3{Trans. Amer. Math. Soc. {\bf{A#1}} (#2) #3}
\def\js#1#2#3{J. SIAM {\bf{#1}} (#2) #3}

\renewcommand{\thefootnote}{\fnsymbol{footnote}}

\begin{document}

\begin{center}


{\Large \bf The Symmetric Stable L\'{e}vy Flights and the Feynman Path Integral}\\
[4mm]

\large{Agapitos Hatzinikitas} \\ [5mm]

{\small University of Aegean, \\
School of Sciences, \\
Department of Statistics and Actuarial-Financial Mathematics, \\
Karlovasi, 83200\\
Samos Greece \\
Email: ahatz@aegean.gr}\\ [5mm]

\end{center}

\begin{abstract}
We determine the solution of the fractional spatial diffusion equation in n-dimensional Euclidean 
space for a ``free" particle by computing the corresponding propagator. 
We employ both the Hamiltonian and Lagrangian approaches which produce exact results for the 
case of jumps governed by symmetric stable L\'{e}vy flights.   
\end{abstract}
\newpage
\section{Introduction}
\renewcommand{\thefootnote}{\arabic{footnote}}
\setcounter{footnote}{0}

\par In nature there is a broad variety of systems in which the correlations in space or time 
give rise to \textit{anomalous} transport whose probability density function (\textit{pdf}) is 
\textit{non-Gaussian} and the squared displacement is non-linear in time or diverges. A 
typical example is an asymmetric particle plume which spreads at a rate inconsistent with the 
classical model. We will 
concetrate on the power-law pattern $E X^2_t\sim K^{\alpha} t^{\alpha}$ in the super-diffusion 
case $1<\alpha<2$, which is manifested 
in a diverse number of systems \cite{weiss}. 
\par In a continuous time random walk framework \cite{metzler} if the jump length and the 
waiting time are considered to be 
independent random variables the joint pdf $\psi(x,t)$ can be written as a product of the 
marginal pdf's for each random variable. It can be shown that if the waiting time satisfies 
a Poisson distribution while the jump length follows a symmetric stable L\'{e}vy 
distribution (see Appendix A for the definition) then a Markov process 
(the first moment of the waiting time is finite) is generated and 
satisfies the fractional spatial diffusion equation \cite{compte} 
\beqr
\frac{\partial P(x,t)}{\partial t}=K^{\alpha}\mathcal{D}^{\alpha}_+ P(x,t).
\label{1}
\feqr  
\noi In \rf{1} the real parameter $\alpha$ ranges into the interval $(1,2]$, the generalised 
diffusion constant has dimensions 
$[K^{\alpha}]=\frac{[L]^{\alpha}}{[T]}$ with $[L], [T]$ being the dimensions of length and 
time, and 
$\mathcal{D}^{\alpha}_+ $ is the Weyl operator as defined in the Appendix A. The solution of 
\rf{1} with the sharp initial condition $\lim_{t\rightarrow 0^+}P(x,t)=\delta(x)$ can be 
obtained 
analytically in terms of the H-function \cite{west} (the general definition of 
Fox functions is given in Appendix B)
\beqr
P(x,t; \alpha)=\frac{1}{\alpha |x|}H_{2, 2}^{1, 1}
\left[ \frac{|x|}{(K^{\alpha}t)^{\frac{1}{\alpha}}}
\Biggl|
\begin{array} {cl}(1, \frac{1}{\alpha}), & (1, \frac{1}{2})\\
(1, 1), & (1, \frac{1}{2}) \end{array}\right].
\label{2}
\feqr
\noi Its structure depends exclusively on the underlying geometry of the space. Taking in \rf{2} 
the $\lim_{\alpha\rightarrow 2^-}P(x, t; \alpha)$ the classical Gaussian solution is
recovered by standard theorems of the Fox functions.
\par The present work aims at the construction of the quantum mechanical path integral over 
L\'{e}vy flights. The outline of the paper is as follows. 
\par In \textit{Section 2} the matrix elements of the time evolution operator in real time are 
calculated by inserting complete sets of eigenstates of momentum and position operators. 
Using the property that plane waves remain eigenstates of the Hermitian operator 
$(-\nabla^2)^{\frac{\alpha}{2}}$ and performing the integrations over the intermediate positions 
we end up with an analytic expression for the 
propagator of the ``free" particle in Euclidean space in one- and three-dimensions. The 
propagators are then written in closed form in terms of the Fox functions. We also derive the 
asymptotic behaviour of the transition probability density in one dimension for large and 
short distances. In two- and n-dimensions 
($n>3$) the propagators are written in integral form using the Bessel functions of the first 
kind whose order is half an odd integer.
\par In \textit{Section 3} we write the propagator in a path integral representation in which 
all paths contribute in the quantum mechanical evolution but they are weighted with the complex 
weight $e^{\frac{i}{\hbar}S_M}$, with $S_M$ representing the action of the particle. 
Descritizing and decomposing the paths into classical and quantum fluctuating trajectories and 
using the binomial expansion we 
derive the asymptotic behaviour of the propagator in the long time limit at fixed position. 
Our result 
is then confirmed by employing the steepest descent method to the result extracted from the 
Hamiltonian approach.
\par In \textit{Appendix} we give all the mathematical background needed for understanding 
some notions of the probability theory as well as integral relations which help to 
manipulate the 
calculations of the text.  We also prove a proposition that justifies the use of plane waves as 
eigenfunctions for the operator $(-\nabla^2)^{\frac{\alpha}{2}}$.    
\section{The Hamiltonian approach}
We consider the time independent, one-dimensional Hamiltonian operator
\beqr
\hat{H}(\hat{p}; \alpha)=\left(\frac{\hat{p}^2}{2m}\right)^{\frac{\alpha}{2}}, 
\quad \alpha\in (1,2)
\label{a1}
\feqr
\noi and investigate the matrix elements of the time evolution operator 
$e^{-\frac{\beta}{\hbar}\hat{H}}$, the so-called propagator 
\beqr
P(x_{in}, x_f, \beta; \alpha)=<x_f|e^{-\frac{\beta}{\hbar}\hat{H}}|x_{in}>.
\label{a2}
\feqr
\noi In \rf{a2} $|x_{in}>$ and $<x_f|$ stand for the initial and final states of the particle 
which are eigenstates of the position operator $\hat{x}$
\beqr
\hat{x}|x_{in}>=x_{in} |x_{in}>, \quad <x_f|\hat{x}=x_f <x_f|.
\label{a3}
\feqr
\noi The position and momentum eigenstates form a complete and orthonormal set of states 
\beqr
\int dx |x><x|=I=\int dp |p><p| \non \\
<x|y>=\delta (x-y); \quad <p|p'>=\delta (p-p')
\label{a31}
\feqr
with inner product
\beqr
<x|p>=\frac{1}{2\pi \hbar}e^{\frac{i}{\hbar}\vec{p}\cdot \vec{x}}
\label{a4}
\feqr
\noi where the angle between $\vec{p}$ and $\vec{x}$ is zero or $\pi$. 
We insert in \rf{a2} N complete sets of momentum eigenstates and N-1 complete sets of 
position eigenstates. The time lapse between the initial and final states has been discretized  
according to $\beta=N\epsilon$ with $\epsilon>0$. We then obtain
\beqr
P(x_0, x_N, \beta; \alpha)&=& <x_N|e^{-\frac{N\epsilon}{\hbar}\hat{H}}|x_0> 
= \prod_{k=1}^{N}\int \int dp_k dx_{k-1}<x_k|e^{-\frac{N\epsilon}{\hbar}\hat{H}(\hat{p})}|p_k>
<p_k|x_{k-1}> \non\\
&=& \frac{1}{(2\pi\hbar)^N}\prod_{k=1}^{N}\int \int dp_k dx_{k-1} \, \, 
e^{\frac{i}{\hbar}p_k(x_k -x_{k-1})-\frac{\epsilon}{\hbar}H(p_k)}
\label{a5}
\feqr
\noi with the identifications $|x_{in}>\equiv |x_0>$ and $|x_f>\equiv |x_N>$. In \rf{a5} we have 
used the property that plane waves are eigenfunctions of the 
Hamiltonian with eigenvalues $|p|^{\alpha}$ (see Appendix A) therefore 
\beqr
e^{-\frac{\epsilon}{\hbar}\hat{H}(\hat{p})}|p_k>=e^{-\frac{\epsilon}{\hbar}H(p_k)}|p_k>.
\label{aa1}
\feqr
\noi Performing the integrations over $x_k$'s we obtain
\beqr
P(x_0, x_N, \beta; \alpha)&=& \frac{\hbar^{N-1}}{2\pi \hbar^N} \int_{-\infty}^{\infty}
\prod_{k=1}^{N} \left( dp_k \, \, 
e^{\frac{i}{\hbar}p_k(x_k-x_{k-1})-\frac{\epsilon}{\hbar}H(p_k)}\right) 
\prod_{l=2}^{N}\delta(p_l-p_{l-1}).
\label{a6}
\feqr 
\noi Using the identity
\beqr
\int_{-\infty}^{\infty}\delta(p_k-p_{k-1})e^{-\frac{\epsilon}{\hbar}H(p_{k-1})}
\delta(p_{k-1}-p_{k-2})dp_{k-1}=\delta(p_k-p_{k-2})e^{-\frac{\epsilon}{\hbar}H(p_k)}
\label{a7}
\feqr
\noi and successively performing the integrations over $p_{N-1},\cdots ,p_1$ we arrive at the 
transition probability density
\beqr
P(x_N-x_0, \beta; \alpha)=\frac{1}{2\pi \hbar}\int_{-\infty}^{\infty}
e^{\frac{i}{\hbar}p_N(x_N-x_0)-\frac{\beta}{\hbar}H(p_N)}dp_N
=\frac{1}{2\pi}\int_{-\infty}^{\infty}
e^{ip_N(x_N-x_0)- c\beta (|p_N|^2)^{\frac{\alpha}{2}}}dp_N. 
\label{a8}
\feqr
\noi In \rf{a8} the momentum has been rescalled and the parameter 
$c=\frac{\hbar^{\alpha -1}}{(2m)^{\frac{\alpha}{2}}}$ will play the role of the diffusion 
constant. 
We recognize expression \rf{a8} to be the Fourier transform of the characteristic 
function of the stable law $S_{\alpha}(0, \frac{\beta\hbar^{\alpha -1}}
{(2m)^{\frac{\alpha}{2}}},0)$. 
\par One can prove that \rf{a8} is the solution of the fractional spatial diffusion equation 
\beqr
\frac{\partial P(x,t)}{\partial t}=-c\left(-\nabla^2\right)^{\frac{\alpha}{2}}P(x,t).
\label{a81}
\feqr
\noi To prove this we set for simplicity $x_0=0$, $x_N=x$, $\beta=t$ and partial integrate 
\rf{a8} with respect to time in which case it reveals that
\beqr
\frac{\partial P(x,t)}{\partial t}=-\frac{1}{2\pi}c\int_{-\infty}^{\infty}
e^{ipx}|p|^{\alpha}e^{- c\beta (|p|^2)^{\frac{\alpha}{2}}}dp=
-c\mathcal{F}^{-1}\left[|p|^{\alpha}e^{- c\beta (|p|^2)^{\frac{\alpha}{2}}}\right]
=-c\left(-\nabla^2\right)^{\frac{\alpha}{2}}P(x,t). 
\label{a82}
\feqr
\noi In deriving the second equality in \rf{a82} we used the result \rf{l3} of Appendix A. 
\par One can proceed by writing the $\beta$-dependent  
integrand of \rf{a8} in terms of the H-Fox function as
\beqr
e^{-\frac{\beta}{\hbar}
\left(\frac{p_N^2}{2m}\right)^{\frac{\alpha}{2}}}=\frac{1}{\alpha}
H^{1, 0}_{0, 1}
\left[ \left(\frac{\beta}{\hbar}\right)^{\frac{1}{\alpha}}\frac{|p_N|}{\sqrt{2m}}\Biggl|
(0, \frac{1}{\alpha})\right]
\label{c7}
\feqr 
\noi thus having
\beqr
I(|x_N-x_0|; \beta, \alpha) &=& 
2\int_{0}^{\infty} \cos \left(\frac{1}{\hbar}p_N(x_N-x_0)\right)
e^{-\frac{\beta}{\hbar}H(p_N)}dp_N \non \\
&=& \frac{2\pi \hbar}{\alpha|x_N-x_0|}
H^{1, 1}_{2, 2}
\left[ \frac{\sqrt{2m}}{\hbar}\left(\frac{\hbar}{\beta}\right)^{\frac{1}{\alpha}}
|x_N-x_0| \Biggl|
\begin{array}{cl} (1,\frac{1}{\alpha}) & (1, \frac{1}{2}) \\
(1, 1) & (1, \frac{1}{2})\end{array} \right].
\label{c8}
\feqr
\noi The propagator is then given by
\beqr
P(x_N-x_0, \beta; \alpha)&=& \frac{1}{2\pi\hbar}I(|x_N-x_0|; \beta, \alpha)\non \\
&=& \frac{1}{\alpha|x_N-x_0|}H^{1, 1}_{2, 2}
\left[ \frac{\sqrt{2m}}{\hbar}\left(\frac{\hbar}{\beta}\right)^{\frac{1}{\alpha}}
|x_N-x_0| \Biggl|
\begin{array}{cl} (1,\frac{1}{\alpha}) & (1, \frac{1}{2}) \\
(1, 1) & (1, \frac{1}{2})\end{array} \right]
\label{c9}
\feqr
\noi a result that was already discussed in \cite{metzler} by solving \rf{1} in 
Fourier-Laplace space.
\par The asymptotic behaviour of the propagator for 
$|x_N-x_0|\rightarrow \infty$ or $|x_N-x_0|\rightarrow 0$ at fixed time, can be determined by 
partial integrating \rf{a8} and rescaling the momentum leading to the expression
\beqr
P(x_N-x_0, \beta; \alpha)= \frac{\beta c \alpha}{\pi |x_N-x_0|^{\alpha +1}}
\int_0^{\infty}\xi^{\alpha -1} \sin \xi 
e^{-\beta c \left(\frac{\xi}{|x_N-x_0|}\right)^{\alpha}}d\xi.
\label{c91}
\feqr 
\begin{itemize}
\item Large x expansion \\
The propagator with the help of \rf{c91} is
\beqr 
P(x_N-x_0, \beta; \alpha)\approx \frac{\beta c \alpha}{\pi |x_N-x_0|^{\alpha +1}}
\Gamma(\alpha) \sin(\frac{\pi\alpha}{2}), \quad \alpha<2
\label{c911}
\feqr
\noi which exhibits the known power-law tail of the L\'{e}vy distribution. Due to this 
property, the mean squared displacement diverges $EX^2_t\rightarrow \infty$.
\item Short x expansion\\
Taylor expanding the cosine and changing the integration variable to 
$w=\beta c |p|^{\alpha}$ in \rf{a8} we have 
\beqr
P(x_N-x_0, \beta; \alpha)\approx \frac{1}{\pi\alpha}\frac{1}{(\beta c)^{\frac{1}{\alpha}}}
\Gamma(\frac{1}{\alpha}).
\label{c912}
\feqr 
\end{itemize}
\par The generalization of the propagator in n-dimensions can be achieved by considering the 
Hamiltonian operator
\beqr
\hat{H}(\hat{p}; \alpha)=\left(\frac{\hat{\vec{p}}^2}{2m}\right)^{\frac{\alpha}{2}}.
\label{c92}
\feqr 
\noi In this case the transition probability density is
\beqr
P(|\vec{x}_N-\vec{x}_0|, \beta; \alpha)=\frac{1}{(2\pi \hbar)^n}\int_{-\infty}^{\infty}\cdots 
\int_{-\infty}^{\infty}dV\,\, e^{\frac{i}{\hbar}\vec{p}_N \cdot 
(\vec{x}_N-\vec{x}_0)-\frac{\beta}{\hbar}H(p_N)}
\label{c10}
\feqr 
\noi where $dV=\prod_{i=1}^{n}dp^i_N$ in Cartesian coordinates, 
$H(p_N)=\left(\frac{\delta_{ij}p^i_N p^j_N}{2m}\right)^{\frac{\alpha}{2}}$ and repeated indices
 are summed over. In deriving \rf{c10} we used the proposition of the Appendix A. 
\par It is obvious that 
expression \rf{c10} depends only on the magnitude but not on the 
direction of $\vec{x}_N-\vec{x}_0$. We distinguish the following cases
\begin{enumerate}
\item \textit{Two-dimensions}
\par We adopt plane polar coordinates, choose the first axis of the $\vec{p}_N$-space in 
the direction of $\vec{x}_N-\vec{x}_0$, and find
\beqr
P(|\vec{x}_N-\vec{x}_0|, \beta; \alpha)=\frac{1}{(2\pi \hbar)^2}\int_{0}^{\infty}dp_N \, 
p_N \, e^{-\frac{\beta}{\hbar}H(p_N)}\int_{0}^{2\pi}d\theta \, \, 
e^{\frac{i}{\hbar}p_N|\vec{x}_N-\vec{x}_0|\cos \theta}. 
\label{c11}
\feqr 
\noi But $\int_{0}^{2\pi}d\theta \, \, e^{ic\cos \theta}=2\pi J_0(c)$, where $J_0$ is the 
standard Bessel function defined in Appendix C. Thus
\beqr
P(|\vec{x}_N-\vec{x}_0|, \beta; \alpha)=\frac{1}{2\pi \hbar^2}\int_{0}^{\infty}dp_N \,\,  
p_N \, e^{-\frac{\beta}{\hbar}H(p_N)}J_0(\frac{p_N}{\hbar}|\vec{x}_N-\vec{x}_0|).
\label{c12}
\feqr
\item \textit{Three-dimensions}
\par Adopting spherical coordinates we find
\beqr
P(|\vec{x}_N-\vec{x}_0|, \beta; \alpha)&=&\frac{1}{(2\pi \hbar)^3}\int_{0}^{\infty}
dp_N \,\, p_N^2 \,  
e^{-\frac{\beta}{\hbar}H(p_N)}\int_{-1}^{1}(d\cos \theta) \, \, 
e^{\frac{i}{\hbar}p_N|\vec{x}_N-\vec{x}_0|\cos \theta}\int_0^{2\pi}d\phi\non \\ 
&=& -\frac{1}{2\pi^2 \hbar}\frac{1}{|\vec{x}_N-\vec{x}_0|}
\frac{d}{d|\vec{x}_N-\vec{x}_0|}\int_{0}^{\infty}
dp_N e^{-\frac{\beta}{\hbar}H(p_N)}\cos\left(\frac{p_N}{\hbar}|\vec{x}_N-\vec{x}_0|\right)
\non \\
&=& -\frac{1}{2\pi^2 \hbar}\frac{1}{|\vec{x}_N-\vec{x}_0|} \frac{d}{d|\vec{x}_N-\vec{x}_0|}
\non\\
&& \left[\frac{1}{\alpha|\vec{x}_N-\vec{x}_0|} H^{1, 1}_{2, 2}
\left[ \frac{\sqrt{2m}}{\hbar}\left(\frac{\hbar}{\beta}\right)^{\frac{1}{\alpha}}
|\vec{x}_N-\vec{x}_0| \Biggl|
\begin{array}{cl} (1,\frac{1}{\alpha}) & (1, \frac{1}{2}) \\
(1, 1) & (1, \frac{1}{2})\end{array} \right]\right].
\label{c13}
\feqr
\item \textit{n-dimensions}, $n\geq 3$
\par Let $d\sigma$ denotes the uniform probability measure on the unit sphere 
$S^{n-1}=\{x\in R^n : ||x||^2=1 \}$. Then $dV=\frac{2\pi^{n/2}}{\Gamma(n/2)}dp_N d\sigma$ 
where the constant factor represents the area of the sphere in n-dimensions. The squares 
$\{\sigma_j^2 \}$ have a Dirichlet $Di(\frac{1}{2},\cdots, \frac{1}{2})$ joint distribution
\footnote{The pdf of the Dirichlet distribution with parameters 
$(\alpha_1, \cdots , \alpha_{n+1})$ is
\beqr 
f(x_1, \cdots , x_{n}; \alpha_1, \cdots , \alpha _{n+1})=
\frac{\Gamma(\alpha_1 +\cdots +\alpha_{n+1})}{\Gamma(\alpha_1)\cdots \Gamma(\alpha_{n+1})}
x_1^{\alpha_1 -1}\cdots x_n^{\alpha_n -1}\left(1-\sum_{i=1}^{n}x_i \right)^{\alpha_{n+1}-1}
I_{A_n}(y)\non
\label{r1}
\feqr
\noi where $A_n=\{\vec{x}: x_i\geq 0, \sum_{i=1}^{n}x_i\leq 1 \}$.}, 
so each $\sigma_j$ is distributed as the square root of a $Be(\frac{1}{2},\frac{n-1}{2})$ 
\footnote{The pdf of the beta distribution with parameters $(\alpha, \beta)$ is
$f(x;\alpha, \beta)=\frac{\Gamma(\alpha +\beta)}{\Gamma(\alpha)\Gamma(\beta)}x^{\alpha -1}
(1-x)^{\beta -1}I_{(0,\infty)}(x)$.}  
random variable. Thus the integral becomes
\beqr
P(|\vec{x}_N-\vec{x}_0|, \beta; \alpha)&=& \frac{1}{(2\pi \hbar)^n}
\frac{2\pi^{\frac{n}{2}}}{\Gamma(\frac{n}{2})}\int_{0}^{\infty}\int_{S^{n-1}}dp_N d\sigma \,\,
p_N^{n-1}\cos(\frac{\vec{p}_N\cdot (\vec{x}_N-\vec{x}_0)}{\hbar})e^{-\frac{\beta}{\hbar}H(p_N)}
\non \\
&=& \frac{1}{(2\pi \hbar)^n}\frac{2\pi^{\frac{n}{2}}}{\Gamma(\frac{1}{2})\Gamma(\frac{n-1}{2})}
\int_0^{\infty}dp_N \,\, p_N^{n-1}e^{-\frac{\beta}{\hbar}H(p_N)} 
F(p_N,|\vec{x}_N-\vec{x}_0|,n)
\label{c131}
\feqr 
\noi where
\beqr
F(p_N,|\vec{x}_N-\vec{x}_0|,n)&=& \int_0^1 du \, \, \cos (\frac{p_N|x_N-x_0|}{\hbar}\sqrt{u})
u^{-\frac{1}{2}}(1-u)^{\frac{n-3}{2}}\non \\
&=& \sqrt{\pi} \Gamma(\frac{n-1}{2})\left(\frac{2\hbar}{p_N|x_N-x_0|}\right)^{\frac{n-2}{2}}
J_{\frac{n}{2}-1}(\frac{1}{\hbar}p_N|x_N-x_0|). \non \\
\label{c132}
\feqr
\noi Combining \rf{c131} and \rf{c132} we finally have
\beqr
P(|\vec{x}_N-\vec{x}_0|, \beta; \alpha)&=& \frac{1}{(2\pi\hbar)^{\frac{n}{2}}}
\frac{1}{\hbar|x_N-x_0|^{\frac{n}{2}-1}} G(|x_N-x_0|,n)
\label{c133}
\feqr
\noi where $G(|x_N-x_0|,n)=\int_0^{\infty}dp_N \, \, p_N^{\frac{n}{2}}
e^{-\frac{\beta}{\hbar}H(p_N}J_{\frac{n}{2}-1}(\frac{1}{\hbar}p_N|x_N-x_0|)$.
\end{enumerate}
\noi An alternative approach following similar steps as for two and 
three-dimensions is developed in Appendix C. Note that \rf{c133} 
for $\alpha\rightarrow 2^-$ reduces to the multivariable Brownian 
motion with propagator
\beqr
\lim_{\alpha \rightarrow 2^-}P(|\vec{x}_N-\vec{x}_0|, \beta; \alpha)
&=& \left(\frac{m}{2\pi \hbar \beta}\right)^{\frac{n}{2}}
e^{-\frac{m|x_N-x_0|^2}{2\hbar \beta}}.
\label{c17}
\feqr
\par The propagator \rf{c133} is the solution of the spatial fractional diffusion equation
\beqr
\frac{\partial P(\vec{x},t)}{\partial t}=-\frac{\hbar^{\alpha -1}}{(2m)^{\frac{\alpha}{2}}}
(-\nabla^2)^{\frac{\alpha}{2}}P(\vec{x},t)
\label{c18}
\feqr
\noi with the condition $P(\vec{x},t; \vec{x}',t)=\delta(\vec{x}-\vec{x}')$ at equal times 
and $P\rightarrow 0$ as $|\vec{x}-\vec{x}'|\rightarrow \infty$.  
\section{The Langrangian Approach}

\par Consider the action of the one-dimensional ``free" particle in Minkowski space
\beqr
S_M= f(\alpha)\int_{t_i}^{t_f} \left[\left(\dot{x}\right)^2\right]
^{\frac{\alpha}{2(\alpha -1)}} dt
\label{b1}
\feqr
\noi where the function $f(\alpha)$ has dimensions 
$[E]\left(\frac{[T]}{[L]}\right)^{\frac{\alpha}{\alpha -1}}$ 
(with [E], [L], [T] representing the dimensions of energy, lenght and time respectively)
and is given by 
\beqr
f(\alpha)=(\alpha -1)\left( \frac{(2m)^{\frac{\alpha}{2}}}{\alpha^{\alpha}
R_{\alpha}}\right)^{\frac{1}{\alpha -1}}.
\label{b2}
\feqr
\noi In \rf{b2} $R_{\alpha}$ is a constant with dimenions 
$[R_{\alpha}]=[E]^{1-\frac{\alpha}{2}}$. 
The Lagrangian is invariant under time translations. Applying the transformation 
$t\rightarrow t-t_f$ together with a rescaling by $\beta=t_f-t_i$ the action becomes 
\beqr
S_M=f(\alpha)\left(\frac{1}{\beta}\right)^{\frac{1}{\alpha -1}}\int_{-1}^{0}
\left[\left(\dot{x}\right)^2\right]^{\frac{\alpha}{2(\alpha -1)}}dt.
\label{b3}
\feqr
\noi We decompose now the orbits $x(t)$ of the point particle into a classical path 
$x_{cl}(t)$ and the 
deviations $q(t)$ which are referred onwards as the \textit{quantum fluctuations} of the 
particle orbit. The split and the boundary conditions $x(t)$ satisfy are
\beqr
x(t)=x_{cl}(t)+q(t); \quad x(0)=x(-1)=0.
\label{b4}
\feqr
\noi The classical path correponds to the solution of
the Euler-Lagrange equation 
\beqr
\ddot{x}_{cl}\left(\dot{x}_{cl}\right)^{\frac{2-\alpha}{\alpha -1}}=0
\label{b5}
\feqr
\noi with boundary conditions $x_{cl}(0)=x_{in}$ and $x_{cl}(-1)=x_f$. 
The solution of \rf{b5} with the prescribed conditions is
\beqr
x_{cl}(t)=x_{in}+t(x_i-x_f).
\label{b6}
\feqr
\noi The vanishing of the paths $x(t), x_{cl}(t)$ at the endpoints of the time interval 
$[-1,0]$ forces the quantum fluctuations to obey the conditions $q(0)=q(-1)=0$.
Inserting the decomposition \rf{b4} into the action we have
\beqr
S_M=f(\alpha)\left(\frac{1}{\beta}\right)^{\frac{1}{\alpha -1}}\int_{-1}^{0}
\left(\dot{x}_{cl}+\dot{q}\right)^{\frac{\alpha}{\alpha -1}}dt.
\label{b7}
\feqr 
\noi Using the binomial expansion $(a+b)^r=\sum_{k=0}^{\infty}\left(\begin{array}{l} 
r\\ k \end{array}\right)a^{r-k}b^k$, with $r$, ($k$) be a positive real (nonnegative integer) 
number, $\left(\begin{array}{l} 
r\\ k \end{array}\right)=\frac{r(r-1)\cdots (r-k+1)}{k!}$ and keeping only terms 
quadratic in $q$'s we obtain
\beqr
S_M=f(\alpha) \left(\frac{1}{\beta}\right)^{\frac{1}{\alpha -1}}\int_{-1}^{0}
\left[ \left(\dot{x}_{cl}\right)^{\frac{\alpha}{\alpha -1}}+
\left(\frac{\alpha}{\alpha -1}\right) \left(\dot{x}_{cl}\right)^{\frac{1}{\alpha -1}}
\dot{q}+ \frac{\alpha}{2(\alpha -1)^2} \left(\dot{x}_{cl}\right)^{\frac{2-\alpha}{\alpha -1}}
\dot{q}^2+ O(\dot{q}^3) \right]dt.
\label{b8}
\feqr
\noi The first term is dominated by the contribution along the classical path and the midterm 
vanishes identically after partial integration with respect to $q(t)$ and using the equation 
of motion \rf{b5}. The action then is written as 
\beqr
S_M=f(\alpha) \left(\frac{1}{\beta}\right)^{\frac{1}{\alpha -1}}
\left[\left(x_i-x_f\right)^{\frac{\alpha}{\alpha -1}}+
\frac{\alpha}{2(\alpha -1)^2} \left(x_i-x_f\right)^{\frac{2-\alpha}{\alpha -1}}
\int_{-1}^{0} \left(\dot{q}\right)^2 \right]dt.
\label{b9}
\feqr
\noi The ``free" particle amplitude in configuration space reads
\beqr
<x_f, t_f|x_i, t_i> \!\! &=& \!\! \int_{(x_i, t_i)\rightarrow (x_f, t_f)}
\mathcal{D}x   \, \, e^{\frac{i}{\hbar}S_M} \non \\
\!\! &=& \!\! e^{\frac{i}{\hbar} f(\alpha) \left(\frac{1}{\beta}\right)^{\frac{1}{\alpha -1}}
\left(x_i-x_f\right)^{\frac{\alpha}{\alpha -1}}}
\int \mathcal{D}q \,\, e^{\frac{i}{\hbar}f(\alpha) 
\left(\frac{1}{\beta}\right)^{\frac{1}{\alpha -1}}
\frac{\alpha}{2(\alpha -1)^2} \left(x_i-x_f\right)^{\frac{2-\alpha}{\alpha -1}}
\int_{-1}^{0} \left(\dot{q}\right)^2 dt}.
\label{b10}
\feqr 
\noi Expression \rf{b10} implies a natural factorization of the amplitude into one factor 
steming from the contribution of the  
classical path and another arising from the quantum fluctuations. 
\par Let $t_0=0>t_1>\cdots > t_N=-1$ be a partition of the time interval $[-1,0]$ by points 
of subdivision $t_0, t_1, \cdots , t_N$. The quantum fluctuations are discretized according to
$q_i=q(t_i)$, $i=0, \cdots ,N$ with $q_0=q_N=0$.
The normalization factor of fluctuations occuring in \rf{b10} will be evaluated 
by writing the functional integral as an infinite product of integrals over the discretized 
fields $q_k$ as follows
\beqr
\int \mathcal{D}q \,\, e^{\frac{i}{2\hbar}g(x_i-x_f; \alpha)
\int_{-1}^{0} \left(\dot{q}\right)^2 dt}= \lim_{N \rightarrow \infty} 
\Biggl[ \left(2\pi \hbar\right)^{-N}
\left(\frac{2\pi \hbar g(x_i-x_f; \alpha)}{i\epsilon}\right)^{\frac{N}{2}} 
\prod_{k=1}^{N-1}\int_{-\infty}^{\infty} \, dq_k \, \, 
e^{\frac{i}{2\hbar}S_N}\Biggr]
\label{b11}
\feqr
\noi where $g(x_i-x_f; \alpha)=f(\alpha)\left(\frac{1}{\beta}\right)^{\frac{1}{\alpha -1}}
\frac{\alpha}{(\alpha -1)^2} \left(x_i-x_f\right)^{\frac{2-\alpha}{\alpha -1}}$ and 
$S_N=\frac{1}{\epsilon}g(x_i-x_f; \alpha) \sum_{k=1}^{N}\left(q_k-q_{k-1}\right)^2$. 
To diagonalize the discretized action $S_N$ we expand  
the quantum fluctuations in Fourier modes  
\footnote{This expansion also appears in the continuum time limit by solving 
the Sturm-Liouville problem:
$\ddot{q}+\lambda q=0; \quad q(0)=q(-1)=0$, with the unique solution 
$q(t)=\sum_{n=1}^{\infty} r_n \sqrt{2} \sin(n\pi t)$.}
\beqr
q_k = \sqrt{\frac{2}{N}}\sum_{m=1}^{N-1}\sin \left(\frac{km\pi}{N}\right)r_m 
= \sum_{m=1}^{N-1}O_k^m r_m; \quad k=1,\cdots, N-1.
\label{b12}
\feqr
\noi Substituting \rf{b12} into the discretized action and employing the ortogonality 
relation 
$\sum_{m=1}^{N-1}O_j^m O_k^m=\delta_{j, k}$ we get a product of independent Gaussian integrals 
which can be calculated giving the result
\beqr
\prod_{k=1}^{N-1}\int_{-\infty}^{\infty} \, dq_k \, \, 
e^{\frac{i}{2\hbar\epsilon}g(x_i-x_f; \alpha) \sum_{k=1}^{N}\left(q_k-q_{k-1}\right)^2
}=\left(\frac{2i \epsilon \pi \hbar }{g(x_i-x_f; \alpha)}\right)^{\frac{N-1}{2}}
\frac{1}{\sqrt{N}}.
\label{b13}
\feqr
\noi Combining \rf{b10}, \rf{b11} and \rf{b13} the amplitude is given by 
\beqr
<x_f, t_f|x_i, t_i> \!\! &=& \!\! 
\frac{1}{\sqrt{2i\pi \hbar}} f(\alpha)^{\frac{1}{2}} 
\frac{1}{\beta^{\frac{1}{2(\alpha -1)}}}
\frac{\sqrt{\alpha}}{(\alpha -1)} \left(x_i -x_f\right)^{\frac{2-\alpha}{2(\alpha -1)}}
e^{\frac{i}{\hbar} 
f(\alpha) \left(\frac{1}{\beta}\right)^{\frac{1}{\alpha -1}}
\left(x_i-x_f\right)^{\frac{\alpha}{\alpha -1}}}.
\label{b14}
\feqr
\noi If we analytically continue $\beta$ from Minkowski to Euclidean space using 
$\beta=e^{-i\frac{\pi}{2}(\alpha -1)}\beta_E$ the Euclidean amplitude becomes
\beqr
<x_f, t_f|x_i, t_i> \!\! &=& \!\! 
\frac{1}{\sqrt{2\pi \hbar}} f(\alpha)^{\frac{1}{2}} 
\frac{1}{\beta_E^{\frac{1}{2(\alpha -1)}}}
\frac{\sqrt{\alpha}}{(\alpha -1)} \left(x_i -x_f\right)^{\frac{2-\alpha}{2(\alpha -1)}}
e^{-\frac{1}{\hbar} 
f(\alpha) \left(\frac{1}{\beta_E}\right)^{\frac{1}{\alpha -1}}
\left(x_i-x_f\right)^{\frac{\alpha}{\alpha -1}}}\!\!.
\label{b15}
\feqr
\par We will show now that this amplitude coincides with the long time limit of 
the propagator \rf{c9} at fixed position. We write it as 
\beqr
P(|x_N-x_0|, \beta; \alpha)= 
\frac{1}{2\pi|x_N-x_0|}\int_{-\infty}^{\infty} e^{iz}e^{-\rho |z|^{\alpha}}dz 
\label{b16}
\feqr
\noi  where $\rho=\frac{\beta \hbar^{\alpha -1}}{(2m)^{\frac{\alpha}{2}}|x_N-x_0|^{\alpha}}$. 
To get the best asymptotic approximation of \rf{b16} with $\beta\rightarrow \infty$ 
we apply the method of steepest descent. For this we write 
\beqr
I(\rho; \alpha)=\int_{\gamma} e^{iz-\rho z^{\alpha}}dz=
\int_{\gamma} e^{h(z)}dz 
\label{b17}
\feqr 
\noi where $\gamma$ is the real axis in the complex $z$-plane. We deform the original contour 
of 
integration onto $\gamma'$ so that it takes a steepest descent path through the saddle point
of $h(z)$ avoiding singularities and branch cuts. Then on this contour, most of the contribution to the integral comes from the region around the saddle point, and possibly the endpoints. 
\par The location of the saddle points of $h(z)$ is determined by setting the first derivative 
of $h(z)$ to zero. Thus we find
\beqr
z_0(\alpha)=\frac{1}{\left( \rho \alpha \right)^{\frac{1}{(\alpha -1)}}}
e^{i\frac{\pi}{2}\frac{1}{(\alpha -1)}}.
\label{b18}
\feqr
\noi Expanding $h(z)$ in a Taylor series around $z_0$ we have
\beqr
h(z)&=& h(z_0)+\frac{1}{2!}(z-z_0)^2 h''(z_0)+ \cdots \\
&=& h(z_0)+\frac{1}{2!}(z-z_0)^2 e^{i\theta }|h''(z_0)|+ \cdots 
\label{b19}
\feqr
\noi where $e^{i\theta}$ is the phase of $h''(z_0)$. The linear part of the new path 
$\gamma'$ is 
\beqr
z-z_0=re^{i\phi}
\label{b20}
\feqr
\noi so that
\beqr
dz=e^{i\phi}dr.
\label{b21}
\feqr
\noi Here $e^{i\phi}$ controls the direction of the path. We choose 
the direction of the path to decrease as rapidly as possible such that the function $h(z)$ along 
this path has the form
\beqr
h(z)= h(z_0)-\frac{1}{2!}r^2|h''(z_0)|+ \cdots 
\label{b22}
\feqr
\noi Expression \rf{b22} coincides to \rf{b19} provided that 
\beqr
\phi=\frac{(\pi -\theta)}{2}.
\label{b23}
\feqr
\noi Thus the integral $I(\rho; \alpha)$ becomes 
\beqr
I(\rho; \alpha)= e^{h(z_0)}\int_0^{\infty}e^{-\frac{1}{2}|h''(z_0)|r^2}
\left(\frac{dz}{dr}\right)_{z_0} dr.
\label{b24}
\feqr
\noi Substituting in \rf{b24} $h(z_0)=\left(\frac{1}{\rho \alpha}\right)^{\frac{1}{\alpha -1}}
\left(1-\frac{1}{\alpha}\right)e^{i\frac{\pi}{2}\left(\frac{\alpha}{\alpha -1}\right)}$, 
$|h''(z_0)|=(\alpha -1)\left(\rho \alpha \right)^{\frac{1}{\alpha -1}}$, 
$\phi_0 = \frac{\pi}{4}\left(\frac{2-\alpha}{\alpha -1} \right)$ and  
$\alpha=\frac{2(1+2k)}{1+4k}$, with $k\in Z$, we get
\beqr
P(|x_N-x_0|, \beta; \alpha)= \frac{1}{\sqrt{2\pi \hbar}} f(\alpha)^{\frac{1}{2}} 
\frac{1}{\beta_E^{\frac{1}{2(\alpha -1)}}}
\frac{\sqrt{\alpha}}{(\alpha -1)} \left|x_N -x_0\right|^{\frac{2-\alpha}{2(\alpha -1)}}
e^{-\frac{1}{\hbar} 
f(\alpha) \left(\frac{1}{\beta_E}\right)^{\frac{1}{\alpha -1}}
\left|x_N-x_0\right|^{\frac{\alpha}{\alpha -1}}}.
\label{b25}
\feqr 
\noi As a check one can prove that
\beqr
\lim_{\alpha\rightarrow 2^-}P(|x_N-x_0|, \beta; \alpha)=
&=& \left(\frac{m}{2\pi \hbar \beta}\right)^{\frac{1}{2}}
e^{-\frac{m|x_N-x_0|^2}{2\hbar \beta}}
\label{b26}
\feqr
\noi since the binomial expansion becomes a perfect square and hence all higher order 
contributions vanish in this limit.
\section{Conclusion}
In the super-diffusion case $1<\alpha<2$, we have generalized the transition probability 
density of a particle governed by a 
symmetric and stable L\'{e}vy flight in Euclidean $n$-dimensional space. Following the 
Hamiltonian approach the propagator is written in terms of the Fox function and its derivative 
in one- and three-dimensions correspondingly. In $n=2$ and $n>3$ the propagator is given in 
integral form using the Bessel functions. The Lagrangian approach on the other hand provides 
the asymptotic behaviour of the propagator in the long time limit at fixed position. The result 
produced by the Lagrangian method has also been crossed checked using the steepest descent 
applied to the exact result derived from the Hamiltonian approach.   
\addcontentsline{toc}{subsection}{Appendix A }
\section*{Appendix A }
\renewcommand{\theequation}{A.\arabic{equation}}
\setcounter{equation}{0}
\textit{Definition 1} \\
Let \textbf{X} be a random vector on $R^n$ with probabiliy distribution $\mu$ and characteristic function 
$\omega$. If $\mu^m=\mu * \cdots *\mu$ denotes the $m$-fold convolution of $\mu$ with itself 
then we say that \textbf{X} is \textit{infinitely divisible} if for each $m\in N$ there exist 
$\textbf{X}_{1,m}, \textbf{X}_{2,m}, \cdots , \textbf{X}_{m,m}$ mutually independent and 
identically distributed random variables such 
that the sum $\textbf{X}_{1,m}+ \textbf{X}_{2,m}+ \cdots+\textbf{X}_{m,m}$ has the same  
distribution as \textbf{X}. 
\par Hence if $\textbf{X}_{i,m}$ has distribution $\mu_{m}$ and characteristic function 
$\omega_m$ then 
$\mu=\mu_m^m$ and $\omega=\omega_m^m$. The \textit{L\'{e}vy-Khintchine \cite{feller} 
representation} states that a 
probability measure $\mu$ on $R^n$ is infinitely divisible iff we can write the 
characteristc function $\omega(p)=E[e^{i<p,X>}]$ in the form 
$\exp (\psi(p))$ where
\beqr
\psi(p)=i<p,c>-\frac{1}{2}<p,Ap>+\int_{R_0^n}\left(e^{i<p,x>}-1-\frac{i<p,x>}{1+||x||^2}\right)
M(dx)
\label{h1a}
\feqr
\noi where $R^n_0:=R^n / \{0\}$, $c\in R^n$, A is a symmetric nonnegative-definite $n\times n$ 
matrix (called the \textit{Gaussian covariance matrix}) and $M$ is a $\sigma$-finite 
Borel measure on $R_0^n$ (called a \textit{L\'{e}vy measure}) such that 
\beqr
\int_{R_0^n}min \{1,||x||^2 \} M(dx)<\infty.
\label{h1b}
\feqr
\noi If $A=0$ then $\mu$ is said to be purely non-Gaussian. The triplet $[c,A,M]$ is unique 
and will be called the generating triplet of the infinitely divisible distribution $\mu$.
\par A special case of infinitely divisible laws are the stable laws in $n=1$. Suppose that 
$\textbf{X},\textbf{X}_1, \cdots , \textbf{X}_m$ denote mutually independent random variables 
with a common distribution F and $\textbf{S}_m=\sum_{i=1}^{m}\textbf{X}_i$. \\
\textit{Definition 2} \\
The distribution F is stable if for each $m\in N$ there exist constants $c_m>0$ and 
$\gamma_m \in R$ such that
\beqr
\textbf{S}_m\stackrel{d}{=}c_m \textbf{X}+\gamma_m
\label{h1c}
\feqr
\noi and F is not concetrated at one point. F is stable in the strict sense if $\gamma_m=0$. 
\par The symbol $\stackrel{d}{=}$ means that the distributions of $\textbf{S}_m$ and 
$\textbf{X}$ differ by location 
and scale parameters. The norming constants are of the form $c_m=m^{\frac{1}{\alpha}}$ with 
$1<\alpha\leq 2$ and the constant $\alpha$ is called \textit{characteristic exponent} of F or 
\textit{index} of the stable law.
\par If we consider $A=0$ and the L\'{e}vy measure to be of the form $M(r,\infty)=pCr^{-\alpha}$ 
and $M(-\infty, -r)=qCr^{-\alpha}$ where $r,C>0$, $\,p,q\geq 0$, $\,p+q=1$ for some 
$0<\alpha\leq 2$ then $\mu$ has characteristic function given by  
\beqr
\psi(p)=i\gamma p-\Biggl\{
\begin{array}{cl} c|p|^{\alpha}\left[1-ib\, sgn(p)\tan (\frac{\pi \alpha}{2}) \right];
\quad \alpha \neq 1 \\
C\frac{\pi}{2}|p|\left[1+\frac{2}{\pi}\, sgn(p)\ln (|p|)\right]; \quad \alpha=1 \end{array}
\label{h1}
\feqr 
\noi where $b=p-q, c=C\frac{\Gamma(2-\alpha)}{1-\alpha}\cos (\frac{\pi \alpha}{2})$ with 
$-1\leq b\leq 1$, $c\geq 0$ and $\gamma \in R$. 
The parameters $\alpha, b, c, \gamma$ characterize the asymptotic behaviour, the skewness, 
the scale and the location of the peak of the stable distribution respectively. Moreover the 
collection $(\alpha, b, c, \gamma)$ is called the \textit{stable law parameters}. A stable law 
generated 
by $(\alpha, b, c, \gamma)$ is often denoted by $S_{\alpha}(b,c,\gamma)$. In the present work 
we set the degree of asymmetry to zero ($b=0$).    
\par The $\alpha$th order Weyl derivatives on the infinite axis are defined as \cite{samko}
\beqr
\left({}_{-\infty}\mathcal{D}_x^{\alpha} f\right)(x)\!\!\!\!\!&=&\!\!\!\!\!
\left(\mathcal{D}_{+}^{\alpha} f\right)(x) \stackrel{def}{=} 
\frac{d^m}{dx^m}\left( I_+^{1-\{\alpha \}}f(x)\right) 
=\frac{1}{\Gamma(m-\alpha)}\left(\frac{d}{dx}\right)^m
\int_{-\infty}^{x}\frac{f(t)}{(x-t)^{\alpha -m +1}}dt, \non \\
\left({}_{x}\mathcal{D}_{\infty}^{\alpha} f\right)(x)\!\!\!\!\!&=&\!\!\!\!\!
\left(\mathcal{D}^{\alpha}_- f\right)(x)
\stackrel{def}{=} 
(-1)^m \frac{d^m}{dx^m}\left( I_-^{1-\{\alpha \}}f(x)\right) \non \\
&=&\frac{(-1)^m}{\Gamma(m-\alpha)}
\left(\frac{d}{dx}\right)^m
\int_{x}^{\infty}\frac{f(t)}{(t-x)^{\alpha -m +1}}dt,  
\label{l1}
\feqr
\noi where $\alpha=[\alpha]+\{\alpha\}$ with $[\alpha], \{\alpha\}$ standing for the integral 
and fractional part ($0<\{\alpha\}<1$) of the real number $\alpha$. Also $m=[\alpha]+1$ and   
$\mathcal{D}^{\alpha}_{\pm} f (I^{1-\{\alpha\}}_{\pm})$ are 
the left- and right-handed fractional derivatives (integrals)and $\Gamma$ the Euler's gamma 
function. It 
can be shown that for $f\in C^{\infty}_{0}(\Omega)$, $\Omega \subset R$ the Fourier transforms 
of \rf{l1} satisfy for $0<\alpha\leq 2$
\beqr
\mathcal{F}\left(\mathcal{D}_{\pm}^{\alpha}f(x)\right)=\left(\mp ip\right)^{\alpha}
\mathcal{F}\left(f(x)\right)=\left(\mp ip\right)^{\alpha}\hat{f}(p)
\label{l2}
\feqr  
\noi where $\mathcal{F}\left(f(x)\right)=\hat{f}(p)=\int_{R}e^{ipx}f(x)dx$ and 
$\left(\mp ip\right)^{\alpha}=|p|^{\alpha}e^{\mp i\frac{\pi}{2}\alpha sgn(p)}$. One can 
prove that
\beqr
\mathcal{D}_+^{\frac{\alpha}{2}}\mathcal{D}_-^{\frac{\alpha}{2}}f(x)=
\mathcal{F}^{-1}\left[(-ip)^{\alpha}(ip)^{\alpha}\hat{f}(p) \right]=
\mathcal{F}^{-1}\left[|p|^{\alpha}\hat{f}(p)\right]
=\left(-\nabla^2 \right)^{\frac{\alpha}{2}}f(x)
\label{l3}
\feqr
\noi where $\nabla^2$ is the one dimensional Laplacian. \\
\textit{Proposition 1}\\
Let $f(x)=e^{i\vec{p}\cdot \vec{x}}$ where $\vec{p}, \vec{x} \in R^n$. Then 
\beqr
(-\nabla^2)^{\frac{\alpha}{2}}f(x)=(p^2)^{\frac{\alpha}{2}}f(x)
\label{l31}
\feqr
\noi where $p^2=\sum_{i=1}^{n}p_i^2$ and $p_i\in R$. \\
\textit{Proof}\\
Define the translation operator $T_{\vec{h}}$ of a function $f(\vec{x})$ by
\beqr
(T_{\vec{h}} f)(\vec{x})=f(\vec{x}-\vec{h})
\label{l4}
\feqr
\noi and the finite difference of order $l$ of $f(\vec{x})$ with step $\vec{h}$ and center at 
the point $\vec{x}$, $\Delta^l_{\vec{h}}$, by
\beqr
(\Delta^l_{\vec{h}}f)(\vec{x})=(I-T_{\vec{h}})^lf(\vec{x})=\sum_{k=0}^l \left(\begin{array}{l} 
l\\ k \end{array}\right) (-1)^k f(\vec{x}-k\vec{h})
\label{l5}
\feqr
\noi where $I$ is the identity operator. It can be shown that the Fourier transform of the 
hypersingular integral
\beqr
D^{\alpha}f=\int_{R^n} \frac{\Delta^l_{\vec{y}} f (\vec{x})}{|y|^{n+\alpha}}d^n y
\label{l6}
\feqr 
\noi is given by
\beqr
\mathcal{F}(D^{\alpha}f(\vec{x}))=|p|^{\alpha} d_{n,l}(\alpha)
\mathcal{F}(f(\vec{x}))  
\label{l7}
\feqr
\noi where the constant $d_{n,l}(\alpha)$ is
\beqr
d_{n,l}(\alpha)=\int_{R^n} d^n u \frac{(1-e^{i\hat{p}\cdot \vec{u}})^l}{|u|^{n+\alpha}}.
\label{l8}
\feqr
\noi In \rf{l8} $\hat{p}$ is the unit vector in the direction of $\vec{p}$. Using as 
$f(x)=e^{i\vec{p}\cdot \vec{x}}$ we have 
\beqr
\left(-\nabla^2\right)^{\frac{\alpha}{2}}e^{i\vec{p}\cdot \vec{x}}&=& 
\frac{1}{d_{n, l}(\alpha)}D^{\alpha}e^{i\vec{p}\cdot \vec{x}}
\non \\
&=& \frac{1}{d_{n, l}(\alpha)}e^{i\vec{p}\cdot \vec{x}}\int_{R^n} 
d^n y \frac{1}{|y|^{n+\alpha}}\sum_{k=0}^{l}
\left(\begin{array}{l} l\\ k \end{array}\right) (-1)^k e^{-ik\vec{p}\cdot \vec{y}}
\non \\
&=& \frac{1}{d_{n, l}(\alpha)}e^{i\vec{p}\cdot \vec{x}}\int_{R^n} 
d^n y \frac{(1-e^{i\vec{p}\cdot \vec{y}})^l}{|y|^{n+\alpha}}
\non \\
&=& (p^2)^{\frac{\alpha}{2}}e^{i\vec{p}\cdot \vec{x}}.
\label{l9}
\feqr
\addcontentsline{toc}{subsection}{Appendix B }
\section*{Appendix B }
\renewcommand{\theequation}{B.\arabic{equation}}
\setcounter{equation}{0}

The class of Fox or H-functions \cite{fox} comprises a large class of special functions known in 
mathematical physics. It is defined in terms of the Mellin-Barnes path integral 
\beqr
H_{p, q}^{m, n}(z)=H_{p, q}^{m, n}\left[ z \Biggl| 
\begin{array}{cl} (a_1, A_1), (a_2, A_2), \cdots, (a_p, A_p) \\
(b_1, B_1), (b_2, B_2), \cdots, (b_q, B_q)\end{array}\right]=\frac{1}{2\pi i}\int_L 
\chi(s)z^s ds
\label{m1}
\feqr
\noi with the integral density
\beqr
\chi(s)=\frac{\prod_{i_1=1}^{m}\Gamma(b_{i_1}-B_{i_1}s) 
\prod_{i_2=1}^{n}\Gamma(1-a_{i_2}+A_{i_2}s)}
{\prod_{i_1=m+1}^{q}\Gamma(1-b_{i_1}+B_{i_1}s)
\prod_{i_2=n+1}^{p}\Gamma(a_{i_2}-A_{i_2}s)}.
\label{m2}
\feqr
\noi The H-function possesses a number of interesting properties \cite{gupta} which we list 
the ones we employed in our presentation.
\begin{itemize}
\item If $k>0$ then,
\beqr
H_{p, q}^{m, n}(z)=k H_{p, q}^{m, n}\left[ z^k \Biggl| 
\begin{array}{cl} (a_1, A_1), (a_2, A_2), \cdots, (a_p, A_p) \\
(b_1, B_1), (b_2, B_2), \cdots, (b_q, B_q)\end{array}\right].
\label{m3}
\feqr
\item Under Fourier cosine transformation, the H-function transforms as
\beqr
\int_0^{\infty}H_{p, q}^{m, n}\left[x \Biggl| \begin{array}{cl} (a_p, A_p) \\
(b_q, B_q)\end{array}\right] \cos (kx)dx
=\frac{\pi}{k}H_{q+1, p+2}^{n+1, m}\left[k\Biggl|
\begin{array}{cl} (1-b_q,B_q), (1, \frac{1}{2}) \\
(1,1), (1-a_p,A_p), (1,\frac{1}{2})\end{array}\right].
\label{m4}
\feqr 
\item The function $z^be^{-z}$ can be represented in terms of the H-function as 
\beqr
z^be^{-z}=H_{0, 1}^{1, 0}\left[z\Biggl|
\begin{array}{cl} (0,0) \\
(b,1)\end{array}\right].
\label{m5}
\feqr
\end{itemize}
\addcontentsline{toc}{subsection}{Appendix C }
\section*{Appendix C }
\renewcommand{\theequation}{C.\arabic{equation}}
\setcounter{equation}{0}
Using spherical coordinates in \rf{c10} one has
\beqr
P(|\vec{x}_N-\vec{x}_0|, \beta; \alpha)&=&\frac{1}{(2\pi \hbar)^n}\int_{0}^{\infty}dp_N \,\,
p_N^{n-1}  
e^{-\frac{\beta}{\hbar}H(p_N)}F(p_N,|\vec{x}_N-\vec{x}_0|,n) 
\label{c14}
\feqr
\noi where
\beqr
F(p_N,|\vec{x}_N-\vec{x}_0|,n)\!=\! \int_{0}^{2\pi}\!\!\! d\theta_1 
\! \left(\prod_{k=2}^{n-2}
\int_{0}^{\pi}
d\theta_k \sin^{k-1}\theta_k \right)
\! \int_{0}^{\pi}\! d\theta_{n-1}\,\, 
\sin^{n-2}\theta_{n-1} e^{\frac{i}{\hbar}p_N|\vec{x}_N-\vec{x}_0|\cos\theta_{n-1}}.
\label{c15}
\feqr
\noi Performing the integrations over the angles  
the integral representation of the propagator is
\beqr
P(|\vec{x}_N-\vec{x}_0|, \beta; \alpha)= \left(\frac{1}{2\pi\hbar}\right)^{\frac{n}{2}}
\frac{1}{\hbar|x_N-x_0|^{\frac{n}{2}-1}} 
\int_{0}^{\infty}dp_N \, \, p_N^{\frac{n}{2}}e^{-\frac{\beta}{\hbar}H(p_N)}
J_{\frac{n}{2}-1}(\frac{p_N}{\hbar}|\vec{x}_N-\vec{x}_0|). 
\label{c16}
\feqr
\par The calculation of the integrals over the angles is based on the formulas \cite{ryzhik}
\beqr
\int_{0}^{\pi}d\theta \, \, \sin^m\theta &=& \Gamma(\frac{1}{2})\frac{\Gamma(\frac{1}{2}(m+1))}
{\Gamma(\frac{1}{2}(m+2))}, \quad m\geq 0 \non \\
\int_{0}^{\pi}d\theta \, \, \sin^m\theta e^{ic\cos\theta}&=& 
\int_{-1}^{1}dx(1-x^2)^{\frac{m-1}{2}}e^{icx}
=2^{1-\frac{m}{2}}\pi \frac{\Gamma(m)}{\Gamma(\frac{m}{2})}c^{-\frac{m}{2}}J_{\frac{m}{2}}(c)
\non \\
\int_0^{1}(1-x^2)^{\nu -\frac{1}{2}}\cos(ax)dx&=&\frac{\sqrt{\pi}}{2}
\left(\frac{2}{a}\right)^{\nu}\Gamma(\nu+\frac{1}{2})J_{\nu}(a) \non \\
\int_0^{\infty}x^{\mu} e^{-ax^2}J_{\nu}(\beta x)dx&=& 
\frac{\Gamma({\nu+\mu+1}{2}}{\beta a^{\frac{\mu}{2}}\Gamma(\nu+1)}e^{-\frac{\beta^2}{8a}}
M_{\frac{\mu}{2},\frac{\nu}{2}}(\frac{\beta^2}{4a}) \non \\
2^{2x-1}\Gamma(x)\Gamma(x+\frac{1}{2})&=&\sqrt{\pi} \Gamma(2x).
\label{n1}
\feqr
\noi In \rf{n1} $J_{\nu}$ is the Bessel function of the first kind of order $\nu$ defined by
the equations \cite{watson}
\beqr
J_{\nu}(z)&=& \frac{1}{2\pi i} \left(\frac{z}{2}\right)^{\nu}
\int_{-\infty}^{0^+}t^{-\nu -1}e^{t-\frac{z^2}{4t}} dt  
\non \\ &=& \left(\frac{z}{2}\right)^{\nu}\sum_{k=0}^{\infty}(-1)^k 
\frac{z^{2k}}{2^{2k}k! \Gamma(\nu+k+1)}
\label{n3}
\feqr
\noi or in a more suitable form by
\beqr
J_{n+\frac{1}{2}}(z)=(-1)^nz^{n+\frac{1}{2}}\sqrt{\frac{2}{\pi}}\frac{d^n}{(zdz)^n}
\left(\frac{\sin z}{z}\right)
\label{n4}
\feqr
\noi where $n$ is a natural number. For completeness we give the expicit expressions for $J_0$ 
and $J_{\frac{1}{2}}$ used for the propagators in two and three dimensions. They are
\beqr
J_0(z)&=& \sum_{k=0}^{\infty}(-1)^k \frac{z^{2k}}{2^{2k}(k!)^2} \non \\
J_{\frac{1}{2}}(z)&=& \sqrt{\frac{2}{\pi z}}\sin z.
\label{n5}
\feqr

\bibliographystyle{plain}

\begin{thebibliography} {99}

\bibitem{weiss} S. Alexander, J. Bernasconi, W. R. Schneider and R. Orbach, 
\rmp{53}{1981}{175};\\ 
H. Weiss and R. J. Rubin, \acp{52}{1983}{363}; \\
B. J. West and W. Deering, \pr{246}{1994}{1};\\
A.I. Saichev and G.M. Zaslavsky, \chaos{7}{1997}{753};\\ 
M. Raberto, E. Scalas and F. Mainardi, \physica{314}{2002}{749}.

\bibitem{metzler} E. W. Montroll, \js{4}{1956}{241};\\
E. W. Montroll and G. H. Weiss, \jmp{10}{1969}{753};\\
D. Bedeaux, K.Lakatos and K. Shuler, \jmp{12}{1971}{2116};\\
V.M. Kenkre, E.W. Montroll and M.F. Shlesinger, \jsp{9}{1973}{45};\\
J. Klafter, A. Blumen and M. F. Shlesinger, \pra{35}{1987}{3081};\\
R. Metzler and J. Klafter, \pr{339}{2000}{1}.

\bibitem{compte} A. Compte, \pre{53}{1996}{4191};\\
R. Metzler, J. Klafter and I. Sokolov, \pre{58}{1998}{1621}.

\bibitem{west} B. J. West, P. Grigolini, R. Metzler and T.F. Nonnenmacher, 
\pre{55}{1997}{99};\\
S. Jespersen, R. Metzler and H.C. Fogedby, \pre{59}{1999}{2736}.

\bibitem{feller} W. Feller, ``\textit{An Introduction to Probability Theory and Its 
Application}, 2nd ed. John Wiley and Sons, New York, 1971, Vol. II. 

\bibitem{samko} S. G. Samko, A. A. Kilbas and O. I. Marichev, ``\textit{Fractional Integrals and 
Derivatives - Theory and Applications}, Gordon and Breach, New York, 1993;\\
K. S. Miller and B. Ross, ``\textit{An Introduction to the Fractional Calculus and Fractional 
Differential Equations}, John Wiley and Sons, New York, 1993. 

\bibitem{fox} C. Fox, \tams{98}{1961}{395}.

\bibitem{gupta} H. M. Srivastava, K. C. Gupta and S. P. Goyal, 
``\textit{The H-functions of One and Two Variables with Applications}", South Asian 
Publishers, New Delhi, 1982. 

\bibitem{ryzhik} I. S. Gradshteyn and I. M. Ryzhik, ``\textit{Table of Integrals, Series, 
and Products}", Academic Press,6th ed (2000).

\bibitem{watson} E. T. Whittaker and G. N. Watson, ``\textit{A Course of Modern Analysis}", 
Cambridge University Press,4th ed (1984).
  

\end{thebibliography}

\end{document}